\documentclass[aps, prl, 10pt, superscriptaddress,
twocolumn,floatfix, notitlepage, citeautoscript]{revtex4-2}

\usepackage{amsmath}
\usepackage{amsfonts}
\usepackage{amssymb}
\usepackage{graphicx}
\usepackage{color}
\usepackage[dvipsnames]{xcolor}
\usepackage{xspace}
\usepackage{stix}
\usepackage[sort&compress]{natbib}
\usepackage[english]{babel}

\newcommand{\ud}[1]{{#1^{\dagger}}}
\newcommand{\bra}[1]{\left\langle #1\right|}
\newcommand{\ket}[1]{\left| #1\right\rangle}
\newcommand{\pket}[1]{\left|\! \left|#1\right\rangle\!\right\rangle}

\newcommand{\mean}[1]{\langle#1\rangle}

\usepackage[colorlinks, linkcolor=RoyalBlue, citecolor=RoyalBlue,
urlcolor=RoyalBlue]{hyperref}

\setcitestyle{super,open={},close={}}

\begin{document} 

\title{Entanglement in Resonance Fluorescence}
\date{\today}

\author{Juan Camilo {L\'opez~Carre{\~n}o}}
\email{juclopezca@gmail.com}
\affiliation{Institute of Theoretical Physics, University of Warsaw,
  ul. Pasteura 5, 02-093, Warsaw, Poland}

\author{Santiago {Berm\'udez~Feijoo}}
\affiliation{Departamento de Física, Universidad Nacional de
  Colombia, Ciudad Universitaria, K.~45 No.~26--85, Bogotá D.C.,
  Colombia}

\author{Magdalena Stobińska}
\affiliation{Faculty of Mathematics, Informatics and Mechanics,
  University of Warsaw, ul. Banacha 2, 02-097 Warsaw, Poland}

\begin{abstract}
  Particle entanglement is a fundamental resource upon which are
  based many quantum technologies. However, the up-to-now best
  sources of entangled photons rely on parametric down-conversion
  processes, which are optimal only at certain frequencies, which
  rarely match the energies of condensed-matter systems that can
  benefit from entanglement.  In this Article, we show a way to
  circumvent this issue, and we introduce a new source of entangled
  photons based on resonance fluorescence delivering photon pairs as
  a superposition of vacuum and the Bell state~$\ket{\Phi^-}$. Our
  proposal relies on the emission from the satellite peaks of a
  two-level system driven by a strong off-resonant laser, whose
  intensity controls the frequencies of the entangled
  photons. Furthermore, the degree of entanglement can be optimized
  for every pair of frequencies, thus demonstrating a clear
  advantage over existing technologies. Finally, we illustrate the
  power of our novel source of entangled single-photon pairs by
  exciting a system of polaritons and showing that they are left in
  a maximally entangled steady state.
\end{abstract}

\maketitle

Resonance fluorescence, the interaction between an artificial atom
and coherent light, has been the subject of fundamental research
from the early stages of quantum optics~\cite{kimble1976,
  allen1987}. In particular, the observation of photon antibunching
from this interaction~\cite{kimble1977a} paved the way for
investigations regarding the quantum character of light, and the
phenomena that has enabled, for instance, the pursue of
single-photon transistors~\cite{chang2007}. In turn, single photon
emission is a key element for many quantum information
technologies~\cite{senellart2017, sinha2019}, allowing the
possibility to design protocols for, e.g., quantum
teleportation~\cite{fattal2004} and quantum
cryptography~\cite{beveratos2002}. Usually, the source of light is a
laser (with a well defined energy) which effectively matches a
single energy transition of the artificial atom, and therefore, in
practice, one deals with the excitation of a so-called ``two-level
system'' (2LS). Thus, the laser can only induce a single excitation
in the atom at the time, and which has led resonance fluorescence to
be regarded as an ultrabright source of quantum
light~\cite{dousse2010, gazzano2012, senellart2017}, with high
single-photon purity~\cite{loredo2016, somaschi2016}. However,
recent investigations that analysed the luminescence with spectral
resolution~\cite{delvalle2012}, have found that the emission form a
2LS actually consists of multiple highly-correlated
photons~\cite{gonzalez-tudela2013, peiris2015, lopezcarreno2017,
  sanchezmunoz2014, diaz-camacho2021, zubizarretacasalengua2022,
  masters2022}.

\begin{figure}[h]
  \includegraphics[width=\linewidth]{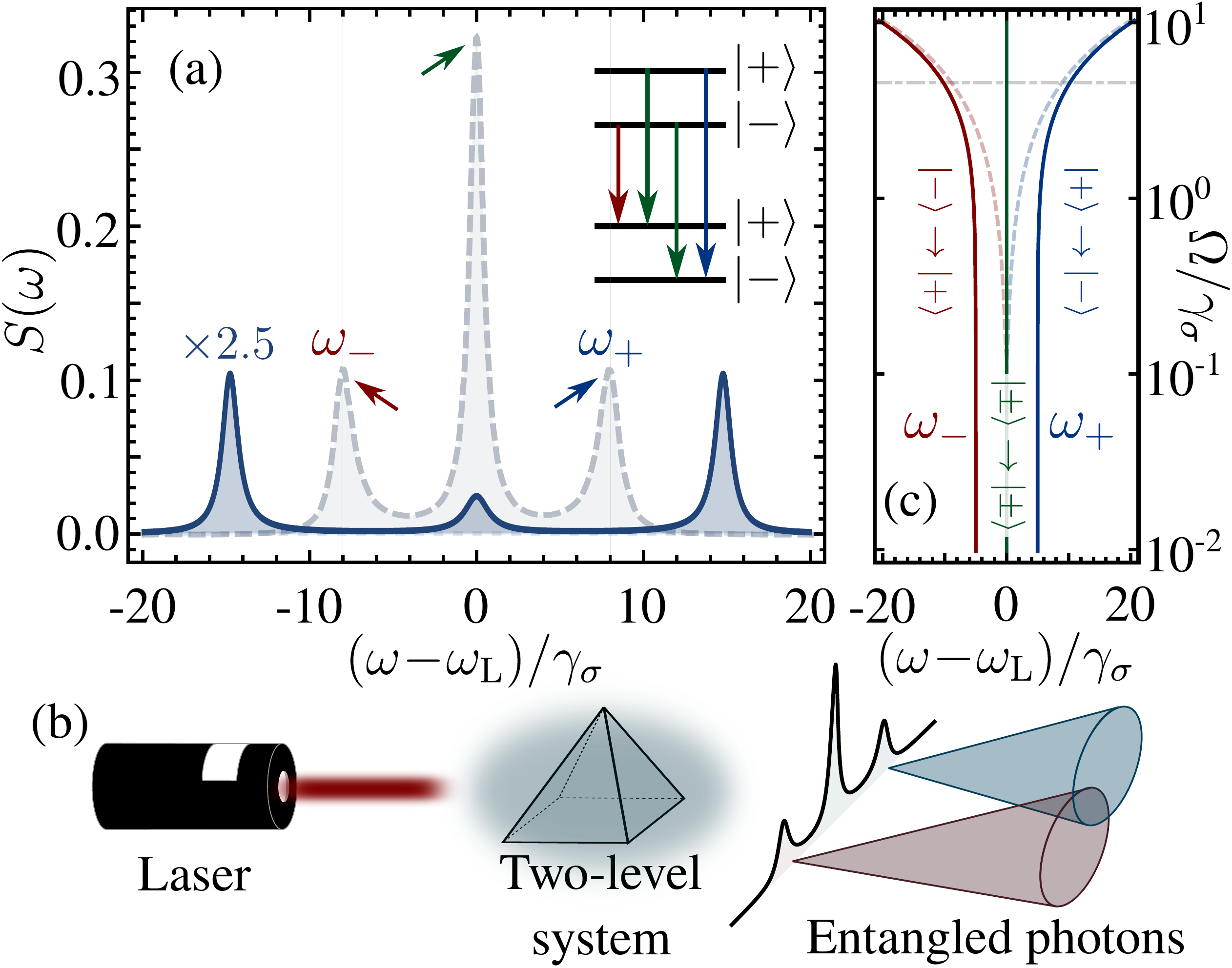}
  \caption{(Color online). The Mollow triplet driven out of
    resonance. (a)~Emission spectrum when the laser is resonant
    (dashed) and detuned (solid) from the 2LS. The emission lines
    are identified with the four possible transitions (two of them
    are degenerate) between the dressed states (shown in the inset).
    (b)~Scheme of our proposed source of entangled photon pairs
    emitted from the sidebands of the Mollow triplet.  (c)~Energy
    transitions enabling the emission from the 2LS as a function of
    the driving intensity. In the detuned case (solid, dark lines),
    and as opposed to the resonant one (dashed, light lines), the
    satellite peaks become the dominant feature of the spectrum, and
    the dynamics is given by transitions that change the quantum
    state of the 2LS, i.e., by transitions of the
    type~$\ket{\pm} \rightarrow \ket{\mp}$ (shown in red and
    blue). For the figure we used~$\gamma_\sigma$ as the unit,
    $\Omega/\gamma_\sigma=4$ (marked in panel (c) as an horizontal
    gray line) and
    $(\omega_\sigma-\omega_\mathrm{L})/\gamma_\sigma = 25/2$.}
  \label{fig:Fri27Aug2021124237CEST}
\end{figure}

Such a multi-photon structure is particularly revealed when the
intensity of the driving laser is strong and the 2LS enters into the
so-called \emph{Mollow regime}~\cite{mollow1969a}, in which the
emission spectrum of the 2LS consists of a triplet, as illustrated
in dashed lines in
Fig.~\ref{fig:Fri27Aug2021124237CEST}(a). Notably, although the
photons emitted from the 2LS are perfectly antibunched, selecting
particular frequency regions of the emission allows to unveil a
richer landscape of photon correlations~\cite{gonzalez-tudela2013,
  peiris2015, lopezcarreno2017}---ranging from antibunching to
superbunching statistics, passing through thermal and uncorrelated
light.  In fact, the statistical variability of the photons emitted
by resonance fluorescence allows to design exotic sources of
light~\cite{sanchezmunoz2014a}, excite other optical
targets~\cite{lopezcarreno2015, lopezcarreno2016a,
  lopezcarreno2016}, and perform the so-called \emph{Mollow
  spectroscopy}~\cite{lopezcarreno2015}, whereby, e.g., the internal
structure of complex and highly-dissipative quantum systems can be
probed with a minimal amount of photons, namely, two.

In this Article, we approach a different type of quantum
correlations, namely entanglement, which has been observed in
Resonance Fluorescence when either a collection of atoms is
considered~\cite{grunwald2010} or its biexciton structure is taken
into account~\cite{delvalle2013} and the 2LS is coupled to a
microcavity~\cite{akopian2006, stevenson2006, dousse2010,
  schumacher2012, johne2008, pathak2009a}. Here, we demonstrate that
time-frequency entanglement can be extracted from resonance
fluorescence, specifically, from the emission of a Mollow triplet,
without the need of coupling it to other objects or taking into
account any additionally internal structures of the 2LS. Instead,
one only needs to include the observation of the emission into the
description.  In fact, we show that when the 2LS is driven out of
resonance, the emission from the sidebands of the triplet behave as
a heralded source of entangled photon pairs, as sketched in
Fig.~\ref{fig:Fri27Aug2021124237CEST}(b).

An important advantage that our proposal has over existing
technologies, e.g., sources based on parametric downconversion
processes is that the entangled photons pairs emitted by our source
follow an antibunched (and not an uncorrelated) statistics. Thus,
while in sources based on SPDC a higher intensity of driving
produces higher-order processes, our source produces more pairs of
entangled photons. Finally, noting that the Mollow regime has been
successfully realized in a variety of systems, including quantum
dots~\cite{konthasinghe2012, ulhaq2012, muller2007a, vamivakas2009,
  flagg2009, ates2009a}, molecules~\cite{wrigge2008}, cold atom
ensembles~\cite{ortiz-gutierrez2019}, confined single
atoms~\cite{ng2022}, photonic chips~\cite{cui2021}, and
superconducting qubits~\cite{astafiev2010, vanloo2013, toyli2016}
(and 2LSs can also be constructed in other platforms including
superconducting circuits~\cite{bouchiat1998,nakamura1999, mooji1999,
  vion2002,martinis2002, koch2007, schreier2008} and photonic
structures~\cite{banyai1993, obrien2009, lodahl2015}), our source is
able to operate on a wide gamut of frequencies and to interface
with, e.g., condensed-matter systems that can benefit from entangled
excitation~\cite{cuevas2018}, thus making resonance fluorescence a
compelling alternative to the existing sources of entangled photons.

The rest of this Article is organized as follows: We first
demonstrate that energy-time entangled between photons emitted from
the sidebands of the 2LS is unveiled simply by including the
observation of the light into the description of our system. Next,
we use a quantum Monte Carlo experiment to demonstrate that, as a
consequence of the detuned excitation, the entangled photon pairs
are heralded. Finally, we show that our source is able to drive
complex condensed-matter systems (e.g.,
exciton-polaritons~\cite{weisbuch1992}) into a maximally entangled
steady state, despite them being immersed within a
highly-dissipative environment.

\begin{figure}[b]
  \includegraphics[width=\linewidth]{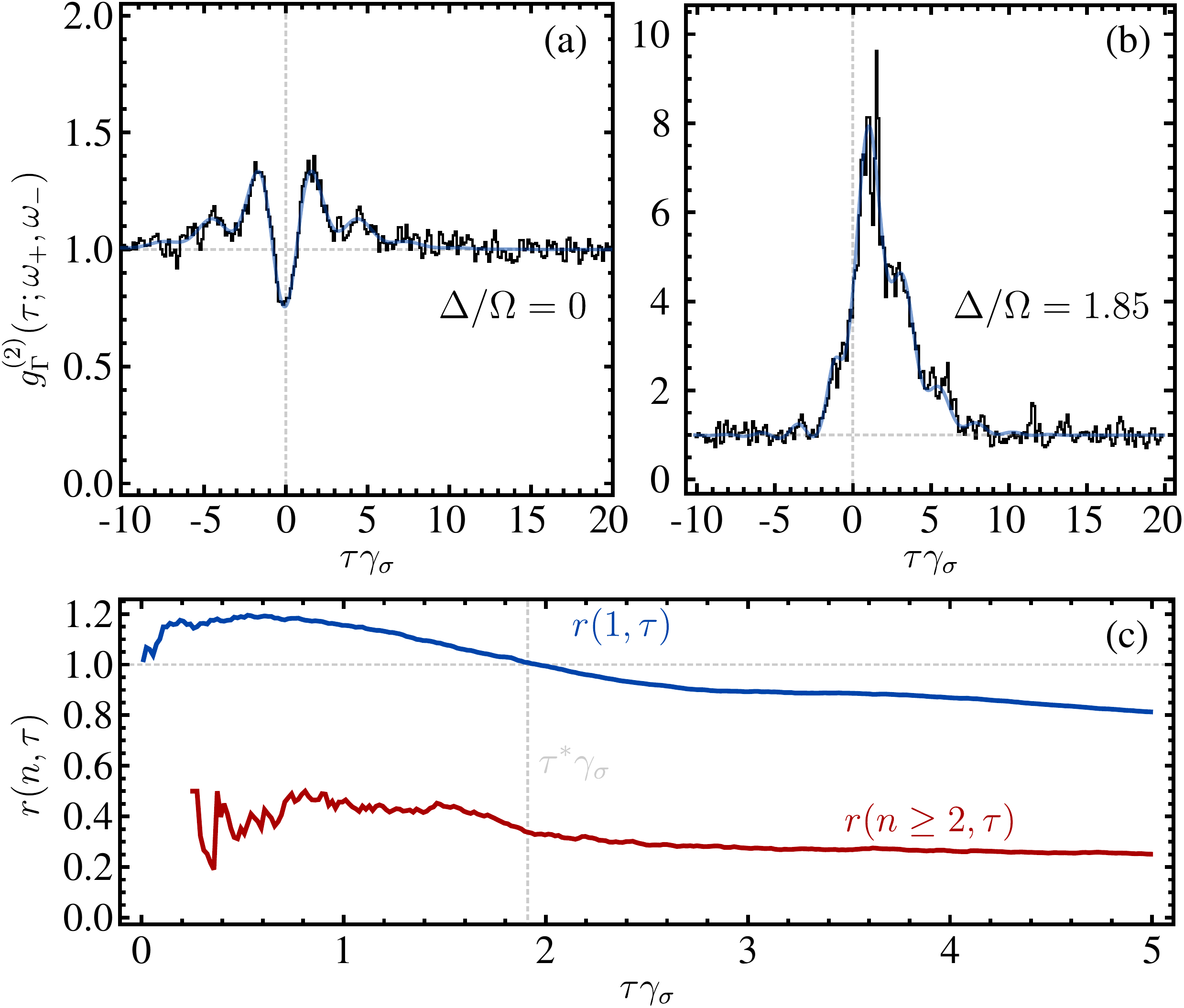}
  \caption{(Color online). Heralded single photons observed through
    a frequency-resolved quantum Monte Carlo experiment. (a,
    b)~Crossed-correlations between photons detected with
    frequencies~$\omega_+$ and~$\omega_-$, showing the agreement
    with theoretical prediction (solid blue lines) and the quantum
    Monte Carlo experiment (black bars). When the driving laser is
    resonant to the 2LS, the correlation function is completely
    symmetric~(cf. panel~a). When the driving is taken out of
    resonance, the shape of the correlation function resembles
    a~$\lambda$, indicating that the emission of a photon with
    frequency~$\omega_+$ heralds the emission of a photon with
    frequency~$\omega_-$ (cf. panel~b).
    (c)~Ratio~$r(n,\tau)=p(n,\tau,\Delta)/p(n,\tau,0)$ of the
    probability to detect one~(blue) and two or more (red) photons
    after detecting the heralding photon, when the driving is made
    out of resonance and in resonance. Taking the laser out of
    resonance enhances the single-photon heralding probability
    within a time-window~$\tau^\ast$ of almost two lifetimes of the
    2LS, as indicated by the dashed vertical line. Additionally, the
    probability to herald two or more photons is suppressed when the
    driving is done out of resonance.  For the figures we
    used~$\gamma_\sigma$ as the unit, and the parameters that
    optimize entangled emission from the sidebands of the triplet
    (cf.~Section II of the Supplemental Material); namely
    $\Omega/\gamma_\sigma=1$, $\Gamma/\gamma_\sigma=1$, and
    $\Delta = 1.85 \Omega$.}
  \label{fig:TueJun7142455CEST2022}
\end{figure} 

\section*{Results}

\begin{figure*}
  \includegraphics[width=\linewidth]{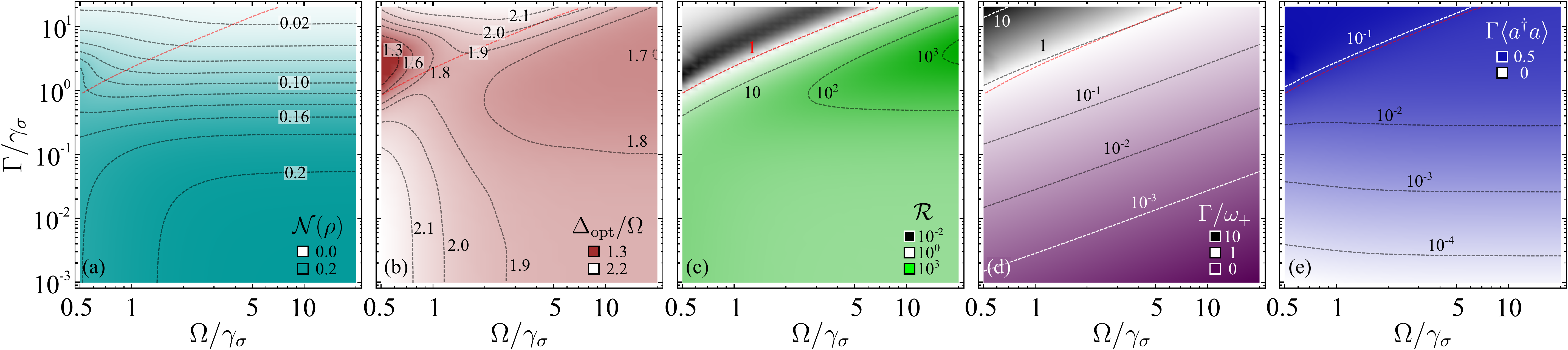}
  \caption{(Color online). Characterisation of our source of
    entangled pairs of single photons. (a)~Maximum logarithmic
    negativity that can be reached as a function of the linewidth of
    the detectors and the intensity of the driving laser, which
    indicates that the optimal condition for extracting entanglement
    is to have narrow detectors observing a Mollow triplet with a
    very large splitting (namely, the bottom right corner of the
    panel).  (b)~The optimum detuning at which the 2LS must be
    driven depends both on the intensity of the driving and the
    linewidth of the detectors or, equivalently, the optical targets
    receiving the entangled photons. (c)~Violation of the CSI
    showing the regions where the negativity obtained in panel~(a)
    is not an artifact of the competition between the detectors for
    the photons of the sidebands. (d)~The ratio between the
    linewidth of the detectors and the splitting between the
    sidebands. It shows that the best configurations are those for
    which this ratio is as small as possible, and that it has to be
    less than 1 to violate the CSI. (e)~Emission rate from the
    detectors, thus completing the mapping of the quality and the
    brightness of the source of entangled single photons based on
    the Mollow triplet.  The red, diagonal, dotted line on each
    panel indicates the boundary above which the emission does not
    violate the CSI.}
  \label{fig:ThuJun2173343CEST2022}
\end{figure*}

\noindent \textbf{Measurement of the photons}\\
\noindent A key aspect of quantum mechanics is that measurements
affect the quantum state of the system under observation. Thus, a
correct description of the emission from the 2LS should also
incorporate in the dynamics the effect of the observation by a
physical detector, namely one able to observe photons of a given
frequency~$\tilde\omega$ with a finite
linewidth~$\tilde\Gamma$. Such a description can be done with the
theory of frequency-resolved correlations~\cite{delvalle2012},
whereby detectors are considered as quantum mechanical objects that
receive the excitation from the source of light without returning
any feedback. In practice, this can be achieved by coupling the
emitter and the detectors with a vanishing
strength~\cite{delvalle2012}, or by using the formalism of
\emph{cascaded systems}~\cite{gardiner1993, carmichael1993}. While
these two approaches yield exactly the same normalized correlations
(i.e., $n^\textrm{th}$-order correlation
functions)~\cite{lopezcarreno2018}, the latter also provides the
correct un-normalized correlations (e.g., the mean population or the
emission rate of the system), and therefore it is the implementation
that we will use throughout this Article. Using the master equation
outlined in the Methods section, we gain access to the correlations
between photons from the Mollow triplet emitted at
frequencies~$\omega_1$ and~$\omega_2$.  In particular, we are
interested in the second-order correlation between photons emitted
at the frequencies of the satellite peaks (with
energies~$\omega_\pm$, c.f.~the scheme in
Fig.~\ref{fig:Fri27Aug2021124237CEST}, associated to bosonic
annihilation operator~$a_1$ and~$a_2$) separated by a time~$\tau$,
i.e., we compute
$g_{12}^{(2)}(\tau)=\mean{\ud{a_1}\ud{a_2}(\tau)a_2(\tau)a_1}/(
\mean{\ud{a_1}a_1} \mean{\ud{a_2}a_2})$ for the case
where~$\omega_1= \omega_+$ and~$\omega_2=\omega_-$. Note that while
the inclusion of detection into the description of the emission
leads to a loss of antibunching of the
signal~\cite{lopezcarreno2022}, it also unveils the multi-photon
structure behind the dynamics of the
2LS~\cite{lopezcarreno2017}. When the latter is driven resonantly by
a laser, its~$g_{12}^{(2)}(\tau)$ is completely symmetric, as shown
in Fig.~\ref{fig:TueJun7142455CEST2022}(a). Such a shape is an
indication that the emission from the high-energy sideband (with
frequency~$\omega_+$) can occur either before or after the emission
from the low-energy peak (with frequency~$\omega_-$); i.e., there is
not a causal relation between consecutive emissions. Such a symmetry
is broken when the driving laser (with
frequency~$\omega_\mathrm{L}$) is taken out of resonance from the
2LS (with natural frequency~$\omega_\sigma$).
Figure~\ref{fig:TueJun7142455CEST2022}(b) shows the case
for~$\Delta\equiv (\omega_\sigma - \omega_\mathrm{L}) =1.85\Omega$,
where it is clear that the emission from the high-energy sideband
occurs \emph{after} the emission of a photon from the low-energy one
(the reverse situation occurs if the detuning becomes
negative). While the type of correlation shown in
Fig.~\ref{fig:TueJun7142455CEST2022}(b) is indicative of a heralded
emission, it does not necessarily imply that the emission consists
of single photons. To show that this is in fact the case, we
performed a frequency-resolved quantum Monte Carlo
simulation~\cite{lopezcarreno2018} of the emission from the two
sidebands of the Mollow triplet to compare the cases when the 2LS is
driven in and out of resonance. Thus, from the simulations we are
able to obtain the probability~$p(n,\tau,\Delta)$ to detect~$n$
photons of energy~$\omega_+$ within a time-window~$\tau$ after a
photon of energy~$\omega_-$ has been measured, provided that the
detuning between the 2LS and the laser
is~$\Delta$. Figure~\ref{fig:TueJun7142455CEST2022}(c) shows the
ratios~$r(n,\tau)=p(n,\tau,\Delta)/p(n,\tau,0)$, which illustrate
that the detuning enhances the probability to detect one photon by
about 20\% (blue line) while simultaneously decreases the
probability to detect two or more photons (red line). Note that the
ratio~$r(1,\tau)$ becomes less than one for time windows larger
than~$\tau^\ast \approx 2/\gamma_\sigma$, shown as a vertical dashed
line in Fig.~\ref{fig:TueJun7142455CEST2022}(c). Thus, the heralded
photons are more likely to be emitted within the time
window~$\tau\leq\tau^\ast$ when the 2LS is driven out of
resonance. Together, Fig.~\ref{fig:TueJun7142455CEST2022}(b) and~(c)
are the evidence that demonstrates that a 2LS driven out of
resonance by a laser is a source of heralded single photons, whose
frequencies correspond to the energies of the sidebands of the
Mollow triplet.\\

\noindent \textbf{Source of entangled photon pairs}\\
\noindent Now that we have established that the emission from the
satellite peaks of the detuned Mollow triplet is composed of highly
correlated pairs of single photons, and that their emission can be
observed in a heralded fashion, we investigate another type of
quantum correlation: entanglement. It has been theoretically
predicted~\cite{sanchezmunoz2014} and experimentally
observed~\cite{peiris2015} that pairs of photons emitted from the
Mollow spectrum at various frequencies violate the Cauchy-Schwarz
inequality (CSI), which can only happen in systems displaying
entanglement~\cite{guhne2009}. Thus, in our Article we quantify
entanglement through the so-called \emph{logarithmic
  negativity}~$\mathcal{N}(\rho)$~\cite{duan2000, simon2000,
  vidal2002}, which is an entanglement monotone~\cite{plenio2005}
that quantifies the degree to which the partial transposition of the
quantum state violates the criterion of
positivity~\cite{horodecki2009}.

Independently of the detuning between the 2LS and the laser that
takes it into the Mollow regime, there are always pairs of
frequencies~$\tilde \omega_1$ and~$\tilde \omega_2$ for which the
CSI is violated~\cite{sanchezmunoz2014}. However, although the
latter is an indication of entanglement, the logarithmic negativity
only becomes nonzero when the laser is taken out of resonance from
the 2LS.  Figure~\ref{fig:ThuJun2173343CEST2022}(a) shows the
maximum~$\mathcal{N}(\rho)$ that can be extracted from photons
emitted at the sidebands, depending on both the intensity of the
driving~$\Omega$ and the linewidth of the detector~$\Gamma$. Here,
each point is obtained for the optimum
detuning~$\Delta_\mathrm{opt}$ between the laser and the 2LS, which
we display in Fig.~\ref{fig:ThuJun2173343CEST2022}(b). For the
largest part of the figure, the detuning that optimizes the
entanglement is around~$\Delta \sim 2 \Omega$. However, in the upper
left corner of the panel we find a region for which the optimal
condition is found near resonance. However, looking at the map of
the CSI violation, shown in in
Fig.~\ref{fig:ThuJun2173343CEST2022}(c), we find that such a region
is compatible with a classical state, as the ratio~$R$ falls below
one (cf. Sec. II of the Supplemental Material). For visual aid, we
have added the~$R=1$ contour as a red dashed line in all the
panels. Note that the behaviour of the CSI, together with the
artificial disruption in the logarithmic negativity, is a
consequence of the competition of the detectors for photons emitted
from the two sidebands, as quantified by the
ratio~$\Gamma/\omega_+$, which we show in
Fig.~\ref{fig:ThuJun2173343CEST2022}(d): the CSI violation starts
precisely when this ratio becomes strictly less than one, and the
photons emitted from the lateral peaks become distinguishable in
frequency. Finally, in Figure~\ref{fig:ThuJun2173343CEST2022}(e) we
show the emission rate of the detectors, i.e.,
$I=\Gamma\mean{\ud{a}a}$, as a function of the intensity of the
laser and the linewidth of the detectors (the detuning between the
2LS and the laser is taken as in panel~(b)), thus making evident the
interplay between the quantity and the quality of the signal: the
highest degree of entanglement is found when the peaks are very well
separated from each other (i.e,~$\Omega/\gamma_\sigma \gg 1$) and
the linewidth of the observer is narrow (with~$\Gamma/\gamma_\sigma$
as small as possible), which in turn comes with the price that such
a narrow linewidth decreases the emission rate of the source. In
fact, in the configurations realized in the bottom right corner of
the panels of Fig.~\ref{fig:ThuJun2173343CEST2022}, the quantum
state of the pairs of entangled single photons is described---with a
97$\%$ fidelity---as the superposition of the vacuum and the
state~$\tilde\rho \equiv \ket{\Phi^-}\bra{\Phi^-}$. The latter is
given by the Bell state
\begin{equation}
  \ket{\Phi^-} = \frac{1}{\sqrt{2}} \left( \ket{0,0} -
    \ket{1,1} \right)\,,
\label{eq:Mon30Jan2023162031CET}
\end{equation}
with a purity of~$91.6\%$, and its contribution to the full quantum
state ranges from 0 to~0.6$\%$. Such a small contribution indicates
that, although the photons are maximally entangled, one needs to
wait for them to be emitted, as discussed above. Furthermore, we
find that entanglement is spoiled as the linewidth of the detectors
becomes large as compared to the emission lines of the triplet. This
is because wide detectors effectively erase the spectral information
of the photons, and the emission from the sidebands becomes
indistinguishable. However, one can overcome such an issue by
driving the 2LS deeper into the Mollow regime, i.e., increasing the
intensity of the laser, and thus taking the satellite peaks further
away from each other. Notably, the pair of
parameters~$(\Omega\,,\Delta)$ that optimize the logarithmic
negativity between the photons from the sidebands of the triplet do
not optimize the violation of the CSI nor maximize their
second-order correlation function (cf. Section II of the
Supplemental Material).\\

\noindent \textbf{Entangling polaritons}\\
\noindent A direct application of the results presented in the
previous section is the excitation of one of the most ubiquitous
systems in condensed matter physics; namely a pair of coupled
bosonic fields. While the latter can represent a large variety of
quantum systems, in the following we will associate them to the
so-called exciton-polaritons (henceforth, simply \emph{polaritons});
which are pseudo-particles arising from the strong coupling between
a photon and an exciton, either within a semiconductor
microcavity~\cite{weisbuch1992} or on an organic
sample~\cite{plumhof2014, lerario2017}. The dynamics of polaritons
driven by resonance fluorescence is given by the master equation
outlined in the Methods section, and we assume that the polaritons
are in the strong coupling regime, in which the energy states become
dressed and the light emitted is observed at the frequencies of the
lower- and upper-polariton branches (cf.~the full derivation in
Sec. III of the Supplemental Material).
\begin{figure}[b]
  \includegraphics[width=\linewidth]{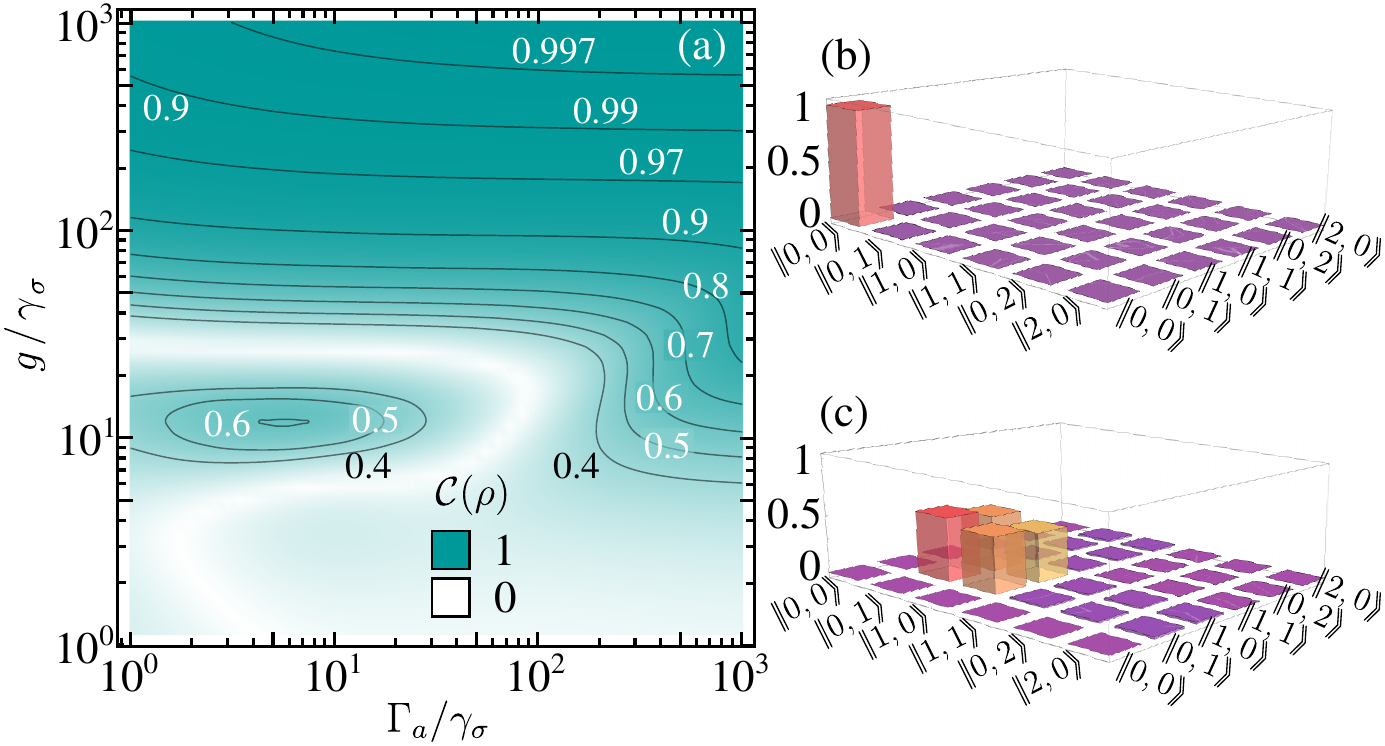}
  \caption{(Color online). Maximally entangled polaritons.
    (a)~Concurrence between the lower- and the upper-polariton
    branches, reaching up to 91\%. (b)~Quantum state tomography of
    the quantum state of polaritons. (c)~Same as panel~(b) after a
    post-selection process removing the vacuum. The remaining
    quantum state has a 0.997 fidelity with a superposition of
    vacuum and the Bell state~$\pket{\Psi^-}$. The figure has been
    done using the parameters that optimize the entanglement of the
    photons emitted by the source,
    namely~$\Omega/\gamma_\sigma=4.9$, $\Delta/\gamma_\sigma = 8.92$
    and letting the photon and the exciton be in resonance with the
    higher- and lower-energy sideband of the triplet,
    respectively. For panels~(b) and~(c) we have also used the
    parameters that yield the maximum concurrence between the
    polaritons, i.e., $\Gamma_a/\gamma_\sigma= 10$,
    $g/\gamma_\sigma\approx 300 $ and~$\Gamma_b\ll \Gamma_a$.}
  \label{fig:Sat10Sep2022114026CEST}
\end{figure}
Figure~\ref{fig:Sat10Sep2022114026CEST}(a) shows the concurrence
between the polariton branches after the vacuum contribution has
been removed through a post-selection process. We find that for
entanglement to be observed it is necessary that: \textit{i)}~the
photonic decay rate has to be at least an order of magnitude larger
than the rate at which the source is emitting light (we find that,
i.e.,~$\Gamma_a/\gamma_\sigma \approx 30$ is typically enough) and
\textit{ii)}~the polariton light-matter coupling should be of the
order of magnitude of the decay of our source. The first condition
guarantees that there are at most two polaritons in the system at
any given time, whereas the latter prevents the excitations to be
localized in a single polariton branch, thus preserving
entanglement. The polariton quantum state that yields the maximum
concurrence is shown in Fig.~\ref{fig:Sat10Sep2022114026CEST}(b)
before and in Fig.~\ref{fig:Sat10Sep2022114026CEST}(c) after a
post-selection process removing the vacuum contribution.  Here, we
find that the quantum state of the polaritons has a 0.997 fidelity
with a superposition between vacuum and the Bell
state~$\pket{\Psi^-}=\frac{1}{\sqrt{2}} \left(\pket{0,1} -
  \pket{1,0} \right)$, where we have used the notation~$\pket{m,n}$
to label a Fock state of polaritons with~$m$ and~$n$ particles in
the lower and upper branch, respectively.  These remarkable results
show the power of our source of entangled photons, while providing
further evidence supporting the observation that polaritons sustain
entanglement~\cite{cuevas2018}, which makes them an attractive
platform to perform, e.g., quantum communication and
cryptography. These applications, however, lay outside the scope of
this Article and will be left for future research.


\section*{Discussion and conclusions}

We have described a novel source of entangled photon pairs, based on
the Mollow triplet regime of resonance fluorescence. In fact, we
showed that when the laser driving the two-level system (2LS)
becomes detuned from the natural frequency of the 2LS, the lateral
peaks of the triplet become the dominant feature of the emission
spectrum. Thus, using the theory of frequency resolved
correlations~\cite{delvalle2012} we showed that, when one focuses on
the photons emitted from such lateral peaks, one finds that their
emission is heralded. Considering the dressed-atom picture, we
showed that these photons, emitted at frequencies~$\omega_+$
and~$\omega_-$ (cf.~the scheme in
Fig.~\ref{fig:Fri27Aug2021124237CEST}), are the result of two atomic
transitions that change the quantum state of the 2LS, and therefore
cannot take place twice in a row. Instead, they alternate in such a
way that the emission of a photon~$\omega_+$ favours the emission of
a photon~$\omega_-$ (note that the opposite order can be achieved if
the laser is blue- instead of red-shifted with respect to the
2LS). We have used a quantum Monte Carlo experiment to show that,
taking the laser out of resonance from the 2LS, not only leads to
the heralding behaviour, but also suppresses emissions consisting of
more than one photons; namely, our system operates as a source of
heralded single photons.

Analysing the quantum correlations between the single photons
emitted from the lateral peaks of the detuned Mollow triplet, we
find that they are emitted in a superposition of vacuum and the Bell
state~$\ket{\Phi^-}$. Furthermore, we have obtained the volume in
the parameter space for which the photons violate the Cauchy-Schwarz
inequality. Thus, we observe that for the photons to display quantum
energy-time entanglement, the linewidth of the detector---or, more
generally, the optical target of the emission---has to be narrower
than the separation between the lateral peaks of the
triplet. Otherwise, the degree of indistinguishability of the
photons emitted from the two peaks decreases, which in turn leads to
the loss of entanglement. Furthermore, we showed the relation
between the emission rate of our source and the degree of
entanglement, as measured through the logarithmic negativity. Thus,
our manuscript can be used as a road map for the experimental
implementation of a source of entangled photons based on resonance
fluorescence.

Lastly, to demonstrate the power of our source, we used it to excite
an ubiquitous quantum system, commonly recurring in condensed-matter
physics, namely a pair of coupled harmonic oscillators embedded in a
dissipative environment. Thus, we use the example of
exciton-polaritons (although our results are also applicable to
systems composed or containing phonons, plasmons, bosonic
nanoparticles, and photonics in general) and showed that our source
is capable to inject entangled particles into the polariton system,
and that, in turn, they are able to maintain such quantum
correlations in spite of the decoherence introduced by spontaneous
decay. We note that our model for polaritons has already been used
to explain experimental results~\cite{cuevas2018}, and while further
elements of decoherence may be relevant in the analysis of the
entanglement in polaritons, their consideration would take us far
away the scope of the present paper and therefore are left for
future references.

\bibliography{Sci-Camilo,Books,arXiv}

\section*{Methods}

\noindent \textbf{Dynamics of the source}\\
\noindent We describe resonance fluorescence as a two-level system
(2LS) with natural frequency~$\omega_\sigma$ driven by a laser with
intensity~$\Omega$. Formally, this is described by the Hamiltonian
(we take~$\hbar=1$ along the paper)
\begin{equation}
  \label{eq:Tue10Aug2021152257CEST}
  H_\sigma = \Delta  \ud{\sigma} \sigma + \Omega(\ud{\sigma} +
  \sigma )\,,  
\end{equation}
where~$\Delta=(\omega_\sigma - \omega_\mathrm{L})$ is the detuning
between the 2LS and the laser of frequency~$\omega_\mathrm{L}$, and
~$\ud{\sigma}$ ($\sigma$) is the creation (annihilation) operator of
the 2LS, which follow the pseudo-spin algebra. The dissipation of
the system is taken into account through the master equation
\begin{equation}
  \label{eq:Tue10Aug2021152528CEST}
  \partial_t \rho = i[\rho, H_\sigma] + \frac{\gamma_\sigma}{2}
  \mathcal{L}_\sigma(\rho)\,, 
\end{equation}
where~$H_\sigma$ is the Hamiltonian in
Eq.~(\ref{eq:Tue10Aug2021152257CEST})
and~$\mathcal{L}_\sigma(\rho) \equiv 2\sigma \rho \ud{\sigma} - \rho
\ud{\sigma}\sigma - \ud{\sigma}\sigma \rho$. When the laser drives
the 2LS with a large intensity, the system enters into the so-called
\emph{Mollow} regime~\cite{mollow1969a}, which is characterized by
its emission spectrum in the shape of a triplet, with a central line
flanked by a pair of symmetric peaks. An archetypal spectrum of a
driven 2LS with~$\Delta=0$ is shown as a dashed line in
Fig.~\ref{fig:Fri27Aug2021124237CEST}(a). The origin of the three
peaks has a natural explanation in the context of the ``dressed
atom'' picture~\cite{cohen-tannoudji1977}, which relates each of the
four possible transitions between consecutive energy manifolds
[shown in the inset of panel (a)] to the emission peaks. Keeping the
intensity of the excitation constant and taking the driving laser
out of resonance from the 2LS, the triplet splits further and the
satellite peaks become the dominant feature of the spectrum as the
central peak loses its intensity. This is shown in solid lines in
Fig.~\ref{fig:Fri27Aug2021124237CEST}(a). The processes that yield
the emission spectrum can be obtained through the diagonalization of
the Liouvillian of the system~\cite{lopezcarreno2016}. Thus,
rewriting Eq.~(\ref{eq:Tue10Aug2021152528CEST})
as~$\partial_t \rho = -M \rho$, one can find the energy of the
transitions as the imaginary part of the eigenvalues of the
matrix~$M$. Figure~\ref{fig:Fri27Aug2021124237CEST}(c) shows in
solid lines the three energy lines available for the emission of the
2LS driven out of resonance. For comparison, we also show in dashed
lines the energies that unfold when the excitation is resonant [and
which give rise to the spectrum shown in dashed lines in panel
(a)]. The main distinction between these two cases is that in the
detuned case the lines are always splitted, even in the limit
when~$\Omega/\gamma_\sigma \rightarrow 0$. In the opposite regime,
when the intensity of the excitation dominates over the detuning,
i.e., when~$\Omega\gg \Delta$, the splitted lines coincide
again. For intermediate intensities, the energy lines are
approximately given
by~$\omega_{\pm} = \omega_\mathrm{L} \pm \sqrt{4\Omega^2 +
  \Delta^2}$ (the exact expression that takes into account the
dissipation is given in Ref.~\cite{lopezcarreno2017}). The
energies~$\omega_{\pm}$ are associated to the
transitions~$\ket{\pm} \rightarrow \ket{\mp}$ (shown in
Fig.~\ref{fig:Fri27Aug2021124237CEST}(c) in blue and red); namely,
quantum jumps that change the quantum state of the 2LS. These types
of transitions dominate the dynamics of the driven 2LS
when~$\Delta \gtrsim \Omega$ and, because they change the quantum
state of the 2LS, the same transition cannot take place twice in a
row. Instead, they take place one after the other and yield a scheme
of photon heralding, as we showed in the results section.\\

\noindent \textbf{Excitation with quantum light}\\
\noindent The theory of cascaded systems~\cite{gardiner1993,
  carmichael1993} allows us to describe the excitation of an optical
target with the emission from our source. Thus, the master equation
describing our system is upgraded to (cf. section I of the
Supplemental Material for details of the derivation)
\begin{multline}
  \label{eq:Tue10Aug2021171632CEST}
  \partial_t \rho = i[\rho,H_\sigma + H_t] +
  \frac{\gamma_\sigma}{2} \mathcal{L}_\sigma(\rho) +
  \frac{\Gamma_1}{2} \mathcal{L}_{a_1}(\rho)+ \frac{\Gamma_2}{2}
  \mathcal{L}_{a_2}(\rho)-{}\\
  {} - \sqrt{\gamma_\sigma \Gamma_1/2} \left \lbrace
    [\ud{a_1},\sigma \rho] + [\rho \ud{\sigma},a_1]
  \right \rbrace - {}\\
  {} - \sqrt{\gamma_\sigma \Gamma_2/2} \left \lbrace
    [\ud{a_2},\sigma \rho] + [\rho \ud{\sigma},a_2]
  \right \rbrace\,,
\end{multline}
where~$H_\sigma$ is the Hamiltonian in
Eq.~(\ref{eq:Tue10Aug2021152257CEST}) and~$H_t$ is the Hamiltonian
describing the internal degrees of freedom of the optical
target. Note that the terms in the bottom two lines of
Eq.~(\ref{eq:Tue10Aug2021171632CEST}) are responsible for the
unidirectional coupling between the source of light and the optical
targets. For the case of the detectors used to measure the photon
correlations in the first part of the Article, we set~$H_t=H_d$ with
the latter defined as
$H_d= (\omega_1-\omega_\mathrm{L}) \ud{a_1}a_1 +
(\omega_2-\omega_\mathrm{L}) \ud{a_2}a_2$, which takes into account
the free energy of the detectors, and we
let~$\Gamma_1 = \Gamma_2=\Gamma$ to be the rate at which the
detectors decay. The terms in the second line of
Eq.~(\ref{eq:Tue10Aug2021171632CEST}) are responsible for the
unidirectional coupling between the source of light and the
detectors. Conversely, for the case of the excitation of polaritons,
we use the convention~$a_1\rightarrow a$, $a_2 \rightarrow b$, and
we replace~$H_t$ with the polariton Hamiltonian
$H_p=(\omega_a-\omega_\mathrm{L})\ud{a}a +
(\omega_b-\omega_\mathrm{L}) \ud{b}b + g(\ud{a}b+\ud{b}a)$, where we
take into account a photon with energy~$\omega_a$ and an exciton
with energy~$\omega_b$ coupled with a strength~$g$. Finally, we
introduced~$\Gamma_1=\Gamma_a$ and~$\Gamma_2=\Gamma_b$, the decay
rates of the photon and exciton modes of the polaritons.

\section*{Data availability}

The data that support the plots within this paper and other findings
of this study are available from the corresponding Author upon
reasonable request.

\section*{Code availability}

The various codes used for modelling the data are available from the
corresponding Author upon reasonable request.

\section*{Acknowledgements}

J.C.L.C. was supported by the Polish National Agency for Academic
Exchange (NAWA) under project TULIP with number PPN/ULM/2020/1/00235
and from the Polish National Science Center (NCN) ``Sonatina''
project CARAMEL with number 2021/40/C/ST2/00155. M.S. was supported
by the European Union’s Horizon 2020 research and innovation
programme under the Marie Skłodowska-Curie project ``AppQInfo''
No. 956071, the National Science Centre ``Sonata Bis'' project
No. 2019/34/E/ST2/00273, and the QuantERA II Programme that has
received funding from the European Union’s Horizon 2020 research and
innovation programme under Grant Agreement No 101017733, project
``PhoMemtor'' No. 2021/03/Y/ST2/00177.

\section*{Author contributions}

J.C.L.C. proposed the idea. J.C.L.C. and S.B.F. developed the
theoretical formalism and the conceptual tools, and then implemented
the theoretical methods and analysed the data. J.C.L.C. and
M.S. contributed material, analysis tools and
expertise. J.C.L.C. wrote the main paper, the Supplemental
Material, and supervised the research. All authors discussed the
results and its implications and commented on the manuscript.

\section*{Competing  interests}

The authors declare no competing interests. 

\end{document}


\title{Supplemental Material: Entanglement in Resonance
  Fluorescence} 
\date{\today}

\author{Juan Camilo {L\'opez~Carre{\~n}o}}
\email{juclopezca@gmail.com}
\affiliation{Institute of Theoretical Physics, University of Warsaw,
  ul. Pasteura 5, 02-093, Warsaw, Poland }

\author{Santiago {Berm\'udez~Feijoo}} \affiliation{Departamento de
  Física, Universidad Nacional de Colombia, Ciudad Universitaria,
  K.~45 No.~26--85, Bogotá D.C., Colombia}

\author{Magdalena Stobińska}
\affiliation{Faculty of Mathematics, Informatics and Mechanics,
  University of Warsaw, ul. Banacha 2, 02-097 Warsaw, Poland}

\maketitle

\section{Measuring the emission: Cascaded Formalism}

In the main text we have described the detection of the emission
from a 2LS with a pair of physical detectors (i.e., observing light
at a fixed frequency~$\tilde\omega$ with a finite
linewidth~$\tilde\Gamma$ and, possibly, non-ideal
efficiency~$\epsilon$) using the so-called ``cascaded
formalism''. In general, such a theory~\cite{gardiner1993,
  carmichael1993} allows to use light with a non-trivial temporal
structure to be used as the source of excitation of an arbitrary
optical target. Physically, the cascaded formalism allows to
describe the coupling the source and the target of the excitation
unidirectionally, i.e., letting the source evolve independently of
the degrees of freedom of the target. In practice, such a coupling
is realised by counteracting a reciprocal coupling (which enters as
a Hamiltonian into the description of the system) with a dissipative
coupling (which is not reciprocal and is taken into account as a
Lindblad term on the master equation). The interplay between these
two couplings yields a negative interference that cancels completely
the terms that bring back from the target to the source of the
light.

In this section we provide the general master equation of a source
of light observed by a single detector (which is the case most
commonly used, and which is textbook material), but we also provide
the generalization to multiple detectors.

\subsection{A single detector}

Consider the excitation of an optical target, to which we associate
an annihilation operator~$a$, by the emission of a quantum optical
source, associated to another annihilation
operator~$\sigma$. Assuming that the source-target system is
described with Hamiltonian~$H$, and the source and target of the
excitation have decay rate~$\gamma_\sigma$ and~$\Gamma$,
respectively, the master equation governing their dynamics is given
by (we use~$\hbar=1$ along the text)
\begin{equation}
  \label{eq:Fri17Sep2021174649CEST}
  \partial_t \rho = i [\rho, H] + \frac{\gamma_\sigma}{2}
  \mathcal{L}_\sigma \rho + \frac{\Gamma}{2} \mathcal{L}_a \rho +
  \sqrt{\epsilon \alpha \gamma_\sigma \Gamma} \left \lbrace [\sigma \rho,
    \ud{a}] + [a,\rho\ud{\sigma}]\right \rbrace\,,
\end{equation}
where~$\mathcal{L}_c \rho = (2c\rho \ud{c} - \rho \ud{c}c -
\ud{c}c\rho)$, $\alpha$ is a factor that guarantees that
Eq.~(\ref{eq:Fri17Sep2021174649CEST}) is physical, and $\epsilon$ is
the detector efficiency (in the following, we let~$\epsilon=1$,
although the non-ideal case can be recover by
re-scaling~$\alpha \rightarrow \epsilon \alpha$). In particular,
note that the master equation~(\ref{eq:Fri17Sep2021174649CEST}) is
\emph{not} evidently written in the Lindblad form. However, one can
define an operator
\begin{equation}
  \label{eq:Fri17Sep2021175340CEST}
  \mathcal{O}_1 = \sqrt{\chi_1\gamma_\sigma} \sigma +
  \sqrt{\chi_2\Gamma} a\,,
\end{equation}
with~$0\leq \chi_1\,,\chi_2 \leq 1$, in such a way that we can write
\begin{equation}
  \label{eq:Fri17Sep2021175533CEST}
  \frac{1}{2} \mathcal{L}_{\mathcal{O}_1}\rho =
  \chi_1\frac{\gamma_\sigma}{2} \mathcal{L}_\sigma \rho +
  \chi_2\frac{\Gamma}{2} \mathcal{L}_a \rho + \frac{1}{2}
  \sqrt{\chi_1\chi_2 \gamma_\sigma \Gamma} \left( 2\sigma
    \rho \ud{a} + 2a \rho \ud{\sigma} - \ud{\sigma} a \rho - \rho
    \ud{\sigma} a - \ud{a}\sigma \rho - \rho \ud{a}\sigma \right)\,.
\end{equation}
The term inside the brackets in the rightmost term of
Eq.~(\ref{eq:Fri17Sep2021175533CEST}) can be rewritten as
\begin{equation}
  \label{eq:Fri17Sep2021180313CEST}
  2 \left( [\sigma\rho, \ud{a}] + [a,\rho\ud{\sigma}]\right) +
  [\rho, \ud{\sigma}a - \ud{a}\sigma]\,,
\end{equation}
which can be replaced back into
Eq.~(\ref{eq:Fri17Sep2021175533CEST}) to yield
\begin{equation}
  \label{eq:Fri17Sep2021180510CEST}
    \frac{1}{2} \mathcal{L}_{\mathcal{O}_1}\rho =
  \chi_1\frac{\gamma_\sigma}{2} \mathcal{L}_\sigma \rho +
  \chi_2\frac{\Gamma}{2} \mathcal{L}_a \rho +
  \sqrt{\chi_1\chi_2 \gamma_\sigma \Gamma} \left \lbrace
    [\sigma\rho, \ud{a}] + [a,\rho\ud{\sigma}] \right \rbrace
 + \frac{1}{2} \sqrt{\chi_1\chi_2 \gamma_\sigma \Gamma}
 [\rho, \ud{\sigma}a - \ud{a}\sigma] \,.
\end{equation}
Reorganising the terms adequately and
introducing~$\alpha \equiv \chi_1\chi_2$, we find that the
dissipative terms in the master
equation~(\ref{eq:Fri17Sep2021174649CEST}) are given by
\begin{equation}
  \label{eq:Fri17Sep2021181005CEST}
  \frac{\gamma_\sigma}{2} \mathcal{L}_\sigma \rho + \frac{\Gamma}{2}
  \mathcal{L}_a \rho + \sqrt{\alpha \gamma_\sigma \Gamma} \left
    \lbrace [\sigma \rho,
    \ud{a}] + [a,\rho\ud{\sigma}]\right \rbrace =
  \frac{1}{2} \mathcal{L}_{\mathcal{O}_1}\rho +
  (1-\chi_1)\frac{\gamma_\sigma}{2} \mathcal{L}_\sigma \rho +
  (1-\chi_2) \frac{\Gamma}{2} \mathcal{L}_a \rho -\frac{1}{2}
  \sqrt{\alpha \gamma_\sigma \Gamma} [\rho, \ud{\sigma}a -
  \ud{a}\sigma]\,, 
\end{equation}
which yields the following master equation
\begin{equation}
  \label{eq:Fri17Sep2021183207CEST}
  \partial_t \rho = i[\rho, H+iH'] +   \frac{1}{2}
  \mathcal{L}_{\mathcal{O}_1}\rho + 
  (1-\chi_1)\frac{\gamma_\sigma}{2} \mathcal{L}_\sigma \rho +
  (1-\chi_2) \frac{\Gamma}{2} \mathcal{L}_a \rho\,.
\end{equation}
Here~$H' =(1/2) \sqrt{\alpha \gamma_\sigma \Gamma} (\ud{\sigma}a -
\ud{a}\sigma)$ is a bi-directional coupling between the source and
the target of the emission. However, the Lindblad term associated to
the operator~$\mathcal{O}_1$ [defined
eq.~(\ref{eq:Fri17Sep2021175340CEST})] provides the destructive
interference that prevents a back-action from the target to the
source of the excitation, thus letting the dynamics of the source
completely independent from the dynamics of the target.

Therefore, the master equation~(\ref{eq:Fri17Sep2021183207CEST})
describes the source-target system (this time, with an explicit term
coupling between the two parties, $H'$), while we identify three
dissipative processes, namely:
\begin{enumerate}
\item a dissipative coupling, associated to the
  operator~$\mathcal{O}_1$, that negatively interferes with the
  Hamiltonian coupling to let the source of the excitation to remain
  oblivious to the dynamics of the target,
\item  the effective decay rate of the source at
rate~$(1-\chi_1) \gamma_\sigma$, and
\item the effective decay rate of the target at
  rate~$(1-\chi_2) \Gamma$.
\end{enumerate}

\subsection{Multiple detectors}

The generalization to multiple targets is required when one analyzes
the emission of the source at two (or more) frequencies. In such a
case, the dynamics of the entire systems is governed by a master
equation similar to Eq.~(\ref{eq:Fri17Sep2021174649CEST}), namely
\begin{equation}
  \label{eq:martes,5deoctubrede2021,125516CEST}
  \partial_t \rho = i[\rho,H] +
  \frac{\gamma_\sigma}{2}\mathcal{L}_\sigma \rho +  \sum_n
  \frac{\Gamma_n}{2} \mathcal{L}_{a_n} \rho + \sum_n \sqrt{\alpha_n
    \gamma_\sigma \Gamma_n} \left \lbrace [\sigma \rho,
    \ud{a_n}] + [a_n,\rho\ud{\sigma}]\right \rbrace\,,
\end{equation}
where we have assumed that each of the optical targets is associated
with a bosonic field, with its corresponding annihilation
operator~$a_n$ and decay rate~$\Gamma_n$.  In analogy to the case of
a single target shown in the previous section, we introduce a set of
jump operators
\begin{equation}
  \label{eq:martes,5deoctubrede2021,133220CEST}
  \mathcal{O}_n = \sqrt{\lambda_n\gamma_\sigma} \sigma +
  \sqrt{(1-\kappa_n)\Gamma_n} a_n\,,
\end{equation}
with~$0\leq \lambda_n, \kappa_n \leq 1$. 

Following the steps shown in Eqs.~(\ref{eq:Fri17Sep2021175533CEST})
and~(\ref{eq:Fri17Sep2021180313CEST}), we can show that the Lindblad
terms associated to these jump operators can be expressed as
\begin{multline}
  \label{eq:martes,5deoctubrede2021,133625CEST}
  \sum_n \frac{1}{2} \mathcal{L}_{\mathcal{O}_n} \rho = \frac{\gamma_\sigma}{2}
  \mathcal{L}\sigma \rho \sum_n \lambda_n + \sum_n \frac{\Gamma_n}{2}
  \mathcal{L}_{a_n}\rho - \sum_n \frac{\kappa_n \Gamma_n}{2}
  \mathcal{L}_{a_n}\rho + \sum_n
  \sqrt{\lambda_n(1-\kappa_n)\gamma_\sigma \Gamma_n} \left \lbrace
    [\sigma\rho,\ud{a_n}] + [a_n,\rho\ud{\sigma}]  \right\rbrace
  +{}\\ 
  {}+ \frac{1}{2} \sum_n \sqrt{\lambda_n(1-\kappa_n)\gamma_\sigma
    \Gamma_n} [\rho, \ud{a_n}\sigma - \ud{\sigma}a_n]\,.
\end{multline}
Adequately reorganizing the terms and letting~$\alpha_n \equiv
\lambda_n (1-\kappa_n)$, we find that the master
equation~(\ref{eq:martes,5deoctubrede2021,125516CEST}) can be
rewritten as
\begin{equation}
  \label{eq:martes,5deoctubrede2021,134641CEST}
  \partial_t \rho = i\left [\rho , H+i \sum_n H'_n \right] + \left (
    1- \sum_n 
    \lambda_n\right) \frac{\gamma_\sigma}{2} \mathcal{L}_\sigma \rho +
  \sum_n \frac{\kappa_n \Gamma_n}{2} \mathcal{L}_{a_n}\rho +
  \sum_n \frac{1}{2} \mathcal{L}_{\mathcal{O}_n}\,,
\end{equation}
where we have
introduced~$H'_n \equiv (1/2)
\sqrt{\lambda_n(1-\kappa_n)\gamma_\sigma \Gamma_n} (\ud{a_n}\sigma -
\ud{\sigma}a_n)$.  Thus, the dynamics of a source of light
exciting~$n$ optical targets can be decomposed into a Hamiltonian
dynamic---governed by the Hamiltonian~$H+i\sum_n H'_n$---and three
types of dissipative processes, namely
\begin{enumerate}
\item the dissipative coupling between the source and the target of
  the excitation (associated to the operators~$\mathcal{O}_n$),
  which prevents the back-action from the latter to the former,
\item the effective decay rate of the source at rate~$(1-\sum_n
  \lambda_n) \gamma_\sigma$, and
\item the effective decay rate of the targets at rates~$\kappa_n
  \Gamma_n$. 
\end{enumerate}
The master equation that we have used to observe entanglement from
the Mollow triplet~(cf. Eq.~(3) of the main text) is a particular
case of Eq.~(\ref{eq:martes,5deoctubrede2021,134641CEST})
when~$\alpha_1=\alpha_2=1/2$; which can be obtained, e.g., by
letting~$\lambda_1=\lambda_2=1/2$ and~$\kappa_1=\kappa_2=0$, which
in physical terms corresponds to the optimization of both the
dissipative coupling and the amount of light that each detector
receives.

\section{Entanglement and Violation of the Cauchy–Schwarz
  inequality}

\begin{figure*}[b]
  \includegraphics[width=\linewidth]{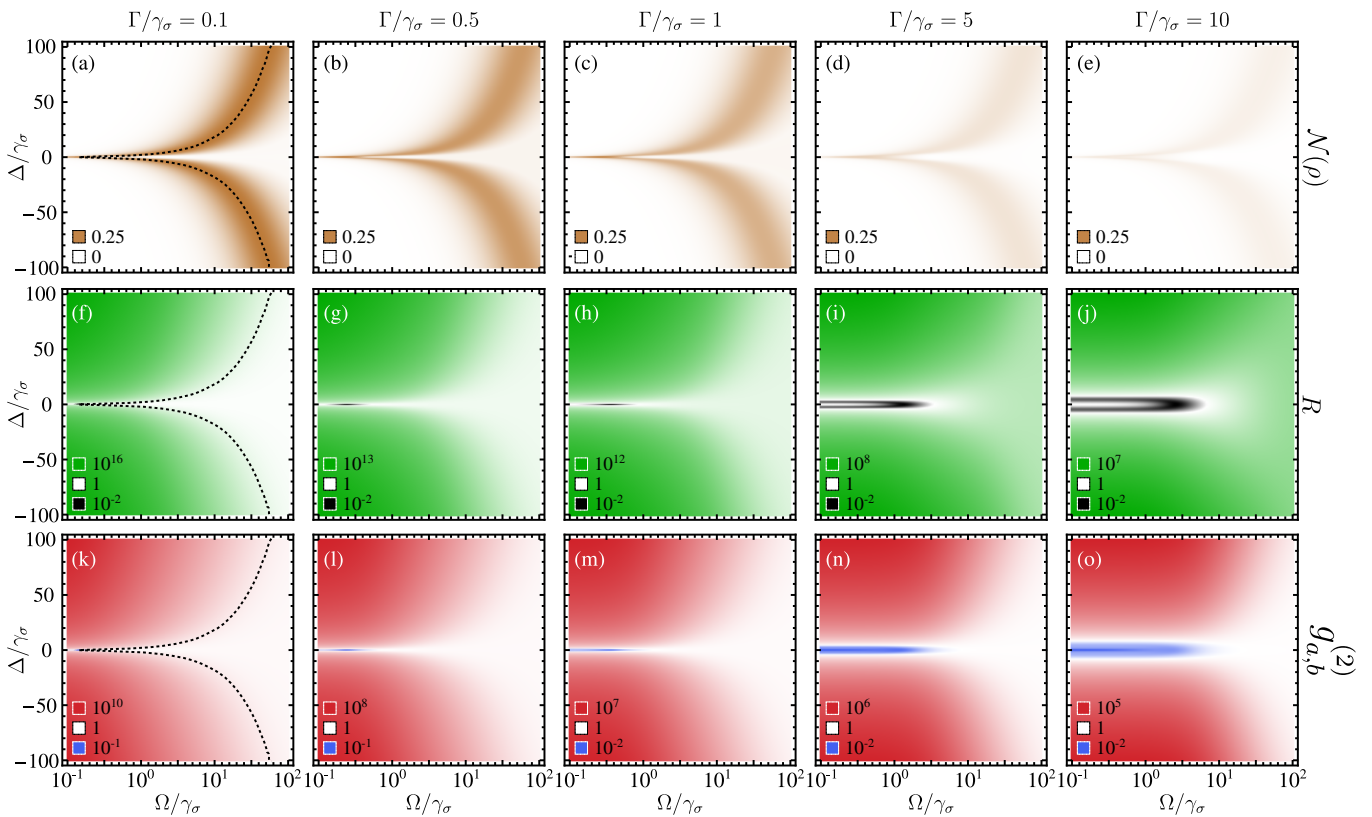}
  \caption{(Color online). Entanglement as shown through the
    logarithmic negativity (upper row), violation of the CSI
    (quantified through the~$R$ coefficient defined in
    Eq.~(\ref{eq:Mon27Jun2022142806CEST}); middle row) and
    second-order correlation function (bottom row) between the
    photons emitted from the sidebands of the 2LS. All the figures
    are shown as a function of the detuning~$\Delta$ between the 2LS
    and the driving laser and the intensity~$\Omega$ with which the
    latter excited the former. From left to right, the columns
    display cases with an increasing linewidth of the detectors,
    showing, in particular, that entanglement is quickly lost as the
    linewidth of the detectors becomes larger than the linewidth of
    the 2LS, i.e., when~$\Gamma\leq \gamma_\sigma$. On the column
    for~$\Gamma/\gamma_\sigma=0.1$ we indicate in dashed lines the
    parameters that provide the maximum entanglement that can be
    distilled from the 2LS. Notably, these parameters do not
    maximize the CSI violation nor the superbunching character of
    the photons. For broader values of~$\Gamma$, the values of the
    maxima decrease, but they are located at the same parameters.}
  \label{fig:Tue28Jun2022165742CEST}
\end{figure*}

The Cauchy-Schwarz inequality (CSI) is a fundamental result of
mathematical analysis, which estates that the inner product between
two vectors, $\vec{u}$ and~$\vec{v}$, cannot be larger than the
product between the norms of each of the vectors, namely
\begin{equation}
  \label{eq:Mon27Jun2022123719CEST}
  |\langle \vec{u}| \vec{v}\rangle|^2 \leq\langle \vec{u} |
  \vec{u}\rangle \cdot \langle \vec{v} | \vec{v}\rangle\,,
\end{equation}
where~$\langle \cdot | \cdot \rangle$ indicates inner product. In
the context of optics, the CSI applies to the intensities and
correlations between fields, and
Eq.~(\ref{eq:Mon27Jun2022123719CEST}) becomes
\begin{equation}
  \label{eq:Mon27Jun2022124515CEST}
  |\mean{I_1 I_2}|^2 \leq \mean{I_1^2}\mean{I_2^2}\,,
\end{equation}
where $I_1$ and~$I_2$ are the intensities of (fluctuating) fields,
and~$\mean{\cdot}$ indicates mean value. While classical states
satisfy Eq.~(\ref{eq:Mon27Jun2022124515CEST}), in quantum mechanics
one can encounter states whose correlations are larger than those
allowed by the CSI~\cite{glauber1963, reid1986}. Therefore, the
violation of the CSI is used as an indicator of nonclassicality.  In
fact, the violation of the CSI has been recently linked with the
appearance of entanglement~\cite{kheruntsyan2012, wasak2014}.

In the main text we deal with the entanglement between photons
emitted from the lateral peaks of the Mollow triplet, namely with
frequencies~$\Omega_\pm$. The observables of these photons are
unveiled by letting them excite a pair of detectors, which have a
finite linewidth, have natural frequencies that match the energy of
the sidebands, and are described with annihilation operators~$a$
and~$b$, both following Bose algebra. Thus, using the formalism of
the second quantization, the
inequality~(\ref{eq:Mon27Jun2022124515CEST}) can be formulated in
terms of the equal-time second-order correlation function of the
operators of the detectors, namely
\begin{equation}
  \label{eq:Mon27Jun2022141845CEST}
  \left[ G_{a,b}^{(2)} \right]^2 \leq G^{(2)}_{a,a} G^{(2)}_{b,b}\,,
\end{equation}
where~$G^{(2)}_{c,d} = \mean{\ud{c}\ud{d}dc}$ for~$c,d \in
\{a,b\}$. Thus, to quantify the degree of violation of the CSI, we
introduce the coefficient
\begin{equation}
  \label{eq:Mon27Jun2022142806CEST}
  R = \frac{\left[ G_{a,b}^{(2)} \right]^2}{G^{(2)}_{a,a}
    G^{(2)}_{b,b}}\,, 
\end{equation}
which is larger than one when the CSI is violated, i.e., when the
state of the detectors is non-classical.
Figure~\ref{fig:Tue28Jun2022165742CEST} shows the logarithmic
negativity (top row), the degree of violation of the CSI (center
row) and the second-order correlation function (bottom row) of the
photons emitted from the sidebands of the Mollow triplet, as a
function of both the intensity of the driving laser~$(\Omega)$ and
its detuning from the 2LS~$(\Delta)$. The various columns
corresponds to an increasing linewidth of the detectors~$\Gamma$. We
find that the three quantities are symmetric with respect to the
detuning of the laser; namely, the behaviour of the emission is the
same regardless of whether the laser is blue- or red-shifted from
the 2LS. Analysing the figure, we observe that although the CSI is
violated for a large range of parameters (cf. the green regions in
the center row), the logarithmic negativity is only nonzero for a
particular set of parameters. In fact,
for~$\Delta \gg \gamma_\sigma$ the driving intensity that optimises
the entanglement is given by~$\Omega \approx 0.58 |\Delta|$, which
is shown as a dashed line in
panel~\ref{fig:Tue28Jun2022165742CEST}(a). Notably, the parameters
that provide the largest entanglement [as measured
through~$\mathcal{N}(\rho)$] do not correspond to a maximum in
either the CSI violation nor the superbunching of the emitted
photons. This means that, although the violation of classical
inequalities is a requisite for entanglement, in our case, an
increase in the nonlocality of the photon pair is not necessarily
accompanied by an increase in the degree of their entanglement.
\begin{SCfigure}[1.2][b]
  \label{fig:Mon4Jul2022162437CEST}
  \includegraphics[width=0.37\linewidth]{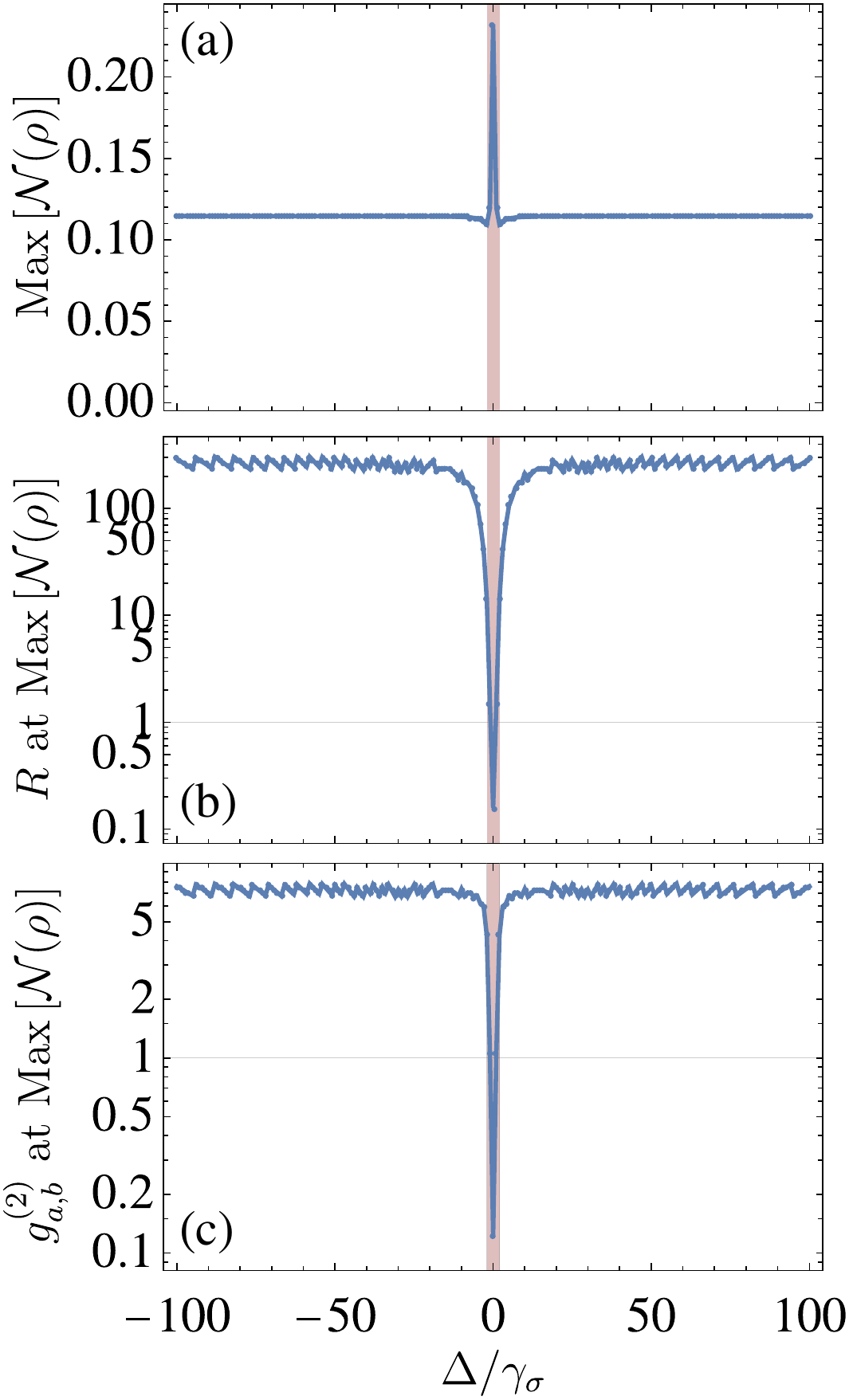}
  \caption{Optimized quantum correlations between photons emitted
    from the sidebands of a 2LS driven coherently with
    detuning~$\Delta$: (a)~Maximum entanglement that can be
    distilled from the photons; (b)~Degree of violation of the
    Cauchy-Schwarz inequality, as defined in
    Eq.~(\ref{eq:Mon27Jun2022142806CEST}); and (c)~Second-order
    correlation between the sidebands for the parameters that
    optimize the entanglement. In the region near resonance
    (when~$\Delta \approx 0$, which we have highlighted in light
    red) we find a surge in the distillable entanglement. However,
    within such a region there coefficient~$R$ falls below 1,
    meaning that the state of the detectors is compatible with a
    classical state. Furthermore, in that same window the emission
    form the sidebands ceases to be heralded, as shown by their
    correlations becoming antibunched. Outside that region, however,
    the emitted photons violate the CSI inequality and are bunched.
    In fact, the three quantities quickly reach a value independent
    of the detuning between the 2LS and the driving laser.}
\end{SCfigure}
Figure~\ref{fig:Mon4Jul2022162437CEST} shows the negativity, the $R$
coefficient and the $g^{(2)}_{a,b}$ for the parameters that optimize
the entanglement, i.e., for the dashed lines in the column
with~$\Gamma/\gamma_\sigma=0.1$ in
Fig.~\ref{fig:Tue28Jun2022165742CEST}. We find that in the region
where~$\Omega_+ \leq \Gamma$ (which we have highlighted as a
vertical red shade on the three panels) the negativity has a spike
while the~$R$ coefficient drops below 1 and the cross-correlation
between the sidebands becomes antibunched. Together, the three
quantities suggest that in this regime the emission is not heralded
(in fact, it is anticorrelated) and the state of the emitted photons
is compatible with a classical state.

This is a consequence of the detection: when the linewidth of the
detectors is large as compared to the size of the Mollow triplet,
the detectors are collecting light from \emph{all} the
frequencies. In particular, they are observing photons emitted from
other energy transitions, which are not strongly correlated with
each others. Such a feature occurs when the linewidth of the
detectors~$\Gamma$ is comparable to the splitting between the
sidebands~$\Omega_+$, which can take place in two distinct
configurations. Firstly, when the triplet is well formed and the
three peaks of the Mollow spectrum can be resolved (i.e.,
when~$\Delta \gg \gamma_\sigma$ and/or~$\Omega\gg
\gamma_\sigma$). In this case, for the detectors to be comparable to
the splitting, it must also be the case
that~$\Gamma\gg \gamma_\sigma$, which implies that we lose the
spectral resolution over the 2LS, and that the observed correlations
are washed out~\cite{gonzalez-tudela2013}. Secondly, when the
emission lines of the triplet are close together so that they cannot
be resolved (i.e., when both~$\Delta \lesssim \gamma_\sigma$
and~$\Omega \lesssim \gamma_\sigma$) and~$\Gamma$ is of the order of
magnitude of~$\gamma_\sigma$. In this case, we retain the spectral
resolution, but the observed photons are indistinguisable in
frequency (thus, the state of light is compatible with a classical
state) and their wavefunction becomes factorisable. In fact, in this
configuration the CSI is either completely satisfied or saturated,
and the emission of the observed photons is uncorrelated (cf. the
regions where~$\Delta/\gamma_\sigma \approx 0$
and~$\Omega < \gamma_\sigma$ in the central and bottom rows of
Fig.~\ref{fig:Tue28Jun2022165742CEST}). Thus, the increase of the
logarithmic negativity shown in
Fig.~\ref{fig:Mon4Jul2022162437CEST}(a) is artificial. It is
obtained from the detectors competing for the photons from both
sidebands, and therefore it is describing ``entanglement'' of a
light mode with itself, which does not have applications in quantum
technologies. However, beyond the highlighted region in
Fig.~\ref{fig:Mon4Jul2022162437CEST}, provided
that~$\Omega_+>\Gamma$, the three quantities reach a constant value
that depends only on the ratio~$\Gamma/\gamma_\sigma$; namely, one
can always find the intensity of the laser~$\Omega$ and its detuning
from the 2LS $\Delta$, that yield the maximum possible
entanglement. Furthermore, by comparing all the columns of
Fig.~\ref{fig:Tue28Jun2022165742CEST}, we find that while the
optimal relation between intensity and detuning is independent of
the linewidth of the detectors, an increase of this parameter only
decreases the amount of distillable entanglement.

\section{Concurrence from polaritons}

In the main text we introduced exciton-polaritons (henceforth,
simply polaritons) as the strong coupling between a photon and an
exciton. Assigning bosonic annihilation operators~$a$ and~$b$ to
these modes, respectively, the polariton Hamiltonian becomes
\begin{equation}
  \label{eq:Fri9Sep2022122902CEST}
  H_p=(\omega_a-\omega_\mathrm{L})\ud{a}a +
  (\omega_b-\omega_\mathrm{L}) \ud{b}b + g(\ud{a}b+\ud{b}a)\,,
\end{equation}
which describes a photon with energy~$\omega_a$ and an exciton with
energy~$\omega_b$ coupled with a strength~$g$. In the strong
coupling regime, the energy levels of
Hamiltonian~(\ref{eq:Fri9Sep2022122902CEST}) hybridize, and the
luminescence of the system takes place at the frequencies of the
so-called \emph{dressed states} of the system. Commonly, these
states are known as the upper- and lower-polariton branches, and
they are formally described with annihilation operators~$u$ and~$l$,
respectively. The latter two are related to the operators of the
photon and of the exciton in the following way
\begin{subequations}
      \label{eq:Fri9Sep2022123756CEST}
  \begin{align}
    \label{eq:Fri9Sep2022123756CESTa}
    l &= \frac{1}{\sqrt{2}} \left(1 +
        \frac{\delta}{\sqrt{\delta^2 + 4g^2}} \right)^{1/2} a -
        \frac{1}{\sqrt{2}} \left(1 -
        \frac{\delta}{\sqrt{\delta^2 + 4g^2}} \right)^{1/2} b\,,\\
        \label{eq:Fri9Sep2022123756CESTb}
    u &= \frac{1}{\sqrt{2}} \left(1 -
        \frac{\delta}{\sqrt{\delta^2 + 4g^2}} \right)^{1/2} a +
        \frac{1}{\sqrt{2}} \left(1 +
        \frac{\delta}{\sqrt{\delta^2 + 4g^2}} \right)^{1/2} b\,,
  \end{align}
\end{subequations}
where we have introduced the
notation~$\delta=\omega_b-\omega_a$. Using the transformation in
Eq.~(\ref{eq:Fri9Sep2022123756CEST}) the
Hamiltonian~(\ref{eq:Fri9Sep2022122902CEST})
becomes~$H_p= (\omega_l-\omega_\mathrm{L})\ud{l}l +
(\omega_u-\omega_\mathrm{L}) \ud{u}u$, with the energies of the
polariton branches defined as
\begin{equation}
  \label{eq:Fri9Sep2022124640CEST}
  \omega_l = \frac{1}{2}\left( \omega_a+\omega_b -
  \sqrt{\delta^2+4g^2} \right) \quad \quad \quad
  \mathrm{and} \quad \quad \quad
  \omega_u = \frac{1}{2}\left( \omega_a+\omega_b +
  \sqrt{\delta^2+4g^2}\right)\,.
\end{equation}
Although polaritons are described as a bosonic field, when we excite
them with the source of entangled photons described in the main
text, we can safely assume that there are, at most, two excitations
within the system. This means that one can limit the Hilbert space
of the system and study the polariton entanglement by turning to a
so-called \emph{detection
  matrix}~$\bar{\theta}$~\cite{cuevas2018}. The latter is
constructed by from mean values of the density
matrix~$\rho_\mathrm{ss}$ obtained as a steady-state solution to the
master equation~(5) of the main text, namely
\begin{equation}
  \label{eq:Fri9Sep2022131111CEST}
  \bar{\theta}\equiv \frac{1}{\mathcal{N}}
  \begin{pmatrix}
    \pbra{0,0}\rho_\mathrm{ss}\pket{0,0}&
    \pbra{0,0}\rho_\mathrm{ss}\pket{1,0} & 
    \pbra{0,0}\rho_\mathrm{ss}\pket{0,1} &
    \pbra{0,0}\rho_\mathrm{ss}\pket{1,1} \\
    h.c. &
    \pbra{1,0}\rho_\mathrm{ss}\pket{1,0} &
    \pbra{1,0}\rho_\mathrm{ss}\pket{0,1} &
    \pbra{1,0}\rho_\mathrm{ss}\pket{1,1} \\
    h.c. & h.c. &
    \pbra{0,1}\rho_\mathrm{ss}\pket{0,1} &
    \pbra{0,1}\rho_\mathrm{ss}\pket{1,1} \\
    h.c. & h.c. & h.c. &
    \pbra{1,1}\rho_\mathrm{ss}\pket{1,1}
 \end{pmatrix}\, ,
\end{equation}
where we have introduced the notation~$\pket{m,n}$ to indicate a
quantum state with~$m$ polaritons in the lower branch and~$n$ in the
upper one. In the definition of the detection matrix we have
included the normalization constant~$\mathcal{N}$ which guarantees
that~$\Tr({\bar{\theta}})=1$, and~$h.c.$ indicates the hermitian
conjugate of the matrix element. Then, from the matrix in
Eq.~(\ref{eq:Fri9Sep2022131111CEST}) we obtain the concurrence
as~$\mathcal{C}(\bar{\theta})
\equiv\max(0,\lambda_1-\lambda_2-\lambda_3-\lambda_4)$ where
the~$\lambda_i$ are the eigenvalues in decreasing order of the
matrix~$\sqrt{\sqrt{\bar{\theta}}\tilde\theta\sqrt{\bar \theta}}$,
where~$\tilde\theta\equiv(\sigma_y\otimes\sigma_y)
\bar{\theta}^T(\sigma_y\otimes\sigma_y)$, and $\sigma_y$ is a Pauli
spin matrix.

\bibliography{Sci-Camilo,Books}